\documentclass[aps, prd, 10pt, twocolumn, superscriptaddress,noshowpacs, bibnotes, nofootinbib, preprintnumbers, longbibliography, hyperref,floatfix]{revtex4-1}

\usepackage{amsmath}
\usepackage{amsfonts}
\usepackage{amssymb}
\usepackage{bbold}
\usepackage{epsfig}
\usepackage{graphicx}
\usepackage{bm}
\usepackage{array}
\usepackage{hyperref}
\usepackage{listings}
\usepackage{color}
\usepackage{float}
\usepackage[normalem]{ulem}

\usepackage{tikz,xcolor,hyperref}

\definecolor{lime}{HTML}{A6CE39}
\DeclareRobustCommand{\orcidicon}{\hspace{-1mm}
	\begin{tikzpicture}
	\draw[lime, fill=lime] (0,0) 
	circle [radius=0.16] 
	node[white] {{\fontfamily{qag}\selectfont \tiny \,ID}};
	\draw[white, fill=white] (-0.0525,0.095) 
	circle [radius=0.007];
	\end{tikzpicture}
	\hspace{-3mm}
}

\foreach \x in {A, ..., Z}{\expandafter\xdef\csname orcid\x\endcsname{\noexpand\href{https://orcid.org/\csname orcidauthor\x\endcsname}
			{\noexpand\orcidicon}}
}

\begin{document}

\title{Do we have enough evidence to invalidate the mean-field approximation adopted to model collective neutrino oscillations?}

\author{Shashank Shalgar\orcidA{}}
\affiliation{Niels Bohr International Academy \& DARK, Niels Bohr Institute,\\University of Copenhagen, Blegdamsvej 17, 2100 Copenhagen, Denmark}
\author{Irene Tamborra\orcidB{}}
\affiliation{Niels Bohr International Academy \& DARK, Niels Bohr Institute,\\University of Copenhagen, Blegdamsvej 17, 2100 Copenhagen, Denmark}

\date{\today}

\begin{abstract}
Recent body of work points out that the mean-field approximation, widely employed to mimic the neutrino field within a neutrino-dense source, might give different results in terms of flavor evolution with respect to the correspondent many-body treatment. 
In this paper, we investigate whether such conclusions derived within a constrained framework should hold in an astrophysical context. We 
show that the plane waves, commonly adopted in the many-body literature to model the neutrino field, provide results that are crucially different with respect to the ones obtained using wavepackets  of finite size streaming with a non-zero velocity.
The many-body approach intrinsically includes coherent and incoherent scatterings. The mean-field approximation, on the other hand, only takes into account the coherent scattering in the absence of the collision term.  Even if incoherent scatterings are included in the mean-field approach, the nature of the collision term is different from that in the many-body approach. Because of this, if only a finite number of neutrinos is considered, as often assumed, the two approaches naturally lead to different flavor outcomes. 
These differences are further exacerbated by vacuum mixing. 
We conclude that existing many-body literature, based on closed neutrino systems with a finite number of particles, is neither able to rule out nor assess the validity of the mean-field approach adopted to simulate the evolution of the neutrino field in dense astrophysical sources, which are  open systems. 
\end{abstract}

\maketitle

\section{Introduction}
\label{sec:intro}
The evolution of neutrino flavor in dense astrophysical environments has been an active field of research due to its rich phenomenology and conceptual complexity, yet not fully explored~\cite{Duan:2010bg, Tamborra:2020cul}. In dense astrophysical environments, such as core-collapse supernovae and compact binary mergers, neutrinos experience a potential due to other neutrinos~\cite{Duan:2006an,Duan:2006jv,Malkus:2014iqa,Malkus:2012ts,Wu:2015fga, Tian:2017xbr, Vlasenko:2018irq,Shalgar:2017pzd}, in addition to the one due to matter~\cite{1978PhRvD..17.2369W,1985YaFiz..42.1441M,Mikheev:1986if}. 
A key reason for the varied phenomenology is the nonlinearity of flavor evolution due to neutrino-neutrino forward scattering. The Hamiltonian in the quantum kinetic equations depends on the neutrino background, which itself evolves dynamically. One of the key features of neutrino flavor evolution in the presence of a neutrino background is that the flavor evolution of all momentum modes is correlated, often referred to as ``collective neutrino flavor conversion.''

Right from the onset of the investigation of the physics of collective neutrino flavor conversion, there have been several controversies in the field due to the seemingly nontrivial quantum mechanical aspects of flavor evolution. For example, one of the earliest discussions focused on whether the self-interaction Hamiltonian has significant off-diagonal components~\cite{Pantaleone:1994ns, Cline:1994ye, Qian:1994uda, Pantaleone:1994ax}, a key feature required for non-linear feedback in the equations of motion.

More recently, several papers have raised questions regarding the validity of the mean-field approximation adopted to model the neutrino flavor evolution---see Ref.~\cite{Patwardhan:2022mxg} for a recent review and references therein.

When neutrinos undergo flavor conversion in the core of a supernova, they stream as wave packets with a finite width from the decoupling region. Traditionally, this implied that each streaming neutrino was assumed to interact with a mean field resembling the neutrino background. 
The mean-field approach adopted to describe the evolution of the one-particle reduced density matrix is based on a crucial assumption that the two-particle correlation that can develop due to the collision term does not affect the flavor evolution. This assumption of molecular chaos ansatz (or {\it Stosszahlansatz}), is valid when one assumes that the average duration of each scattering process is small [$\mathcal{O}(10^{-21})$~s]~\cite{Kersten:2015kio} with respect to the time scale over which the flavor evolution occurs [$\gtrsim \mathcal{O}(10^{-10})$~s]~\cite{Sigl:1992fn, Froustey:2020mcq}.
While being an approximation, such a framework has been widely adopted since the flavor conversion history depends on the average interaction rate of neutrinos with the background along the streaming field.

In this paper, we argue that many of the apparently different conclusions extrapolated by the investigation of the flavor evolution in the many-body and mean-field approaches rely on  tackling  intrinsically different physical systems.
Hence, any assessment of the validity of the mean-field approach to model the evolution of  the neutrino field within an astrophysical source requires further work. Importantly, the many-body literature mostly relies on closed neutrino systems with a limited number of particles. While the propagation and flavor conversion of neutrinos in a dense astrophysical source should be mimicked by considering an open system with an infinitely large number of particles.
The many-body formalism naturally includes momentum-changing processes (both coherent and incoherent scatterings). On the other hand, in the absence of the collisional term, only the coherent forward scattering terms enter the Hamiltonian in the mean-field approach. 

This paper is organized as follows. Section~\ref{sec6b} provides an overview of the approximations intrinsic to the treatment of  a closed neutrino system, in contrast to an astrophysical neutrino-dense source which should be considered as an open system.
In Sec.~\ref{sec3}, we outline the many-body formalism. In Sec.~\ref{sec4}, we explore the neutrino flavor conversion phenomenology foreseen by the many-body approach by relying on the commonly adopted plane wave approximation for neutrinos within a closed interaction box. Section~\ref{sec5} relaxes the plane-wave approximation and focuses on assessing the impact of the size of the wavepacket on the flavor evolution in the many-body approach. We finally discuss and summarize our findings in Sec.~\ref{sec7}. 
An overview of the implicit assumptions made when deriving the equations of motion within the context of the mean-field theory is provided in Appendix~~\ref{sec2}.

\section{Open vs.~Closed system}
\label{sec6b}

In the literature adopting the many-body approach in the context of neutrino oscillations, a finite number of neutrinos, $\mathcal{O}(10$--$1000)$, is confined within a box of finite volume and  assumed to interact for an infinite amount of time. On the other hand, in  neutrino-dense astrophysical sources such as a core-collapse supernovae, the neutrino density is $\mathcal{O}(10^{28})$~cm$^{-3}$ after decoupling, which is relevant for neutrino collective effects in the  bulb model~\cite{Duan:2006an}, and neutrinos stream from the interaction region at the speed of light. 
In order to mimic the flavor evolution of astrophysical neutrinos within an idealized framework, one could then consider a box of finite size, where an infinitely large number of neutrinos represented by wavepackets of finite size streams in and out of the interaction box, while interacting with the other neutrinos met in the interaction volume. As a consequence,  even if two neutrinos undergo a momentum changing collision, they will not see each other again.

In the literature on spin systems, it is common to consider a closed system of interacting spins. This implies that the spins continue to interact with each other for a long period of time and the quantum entanglement between the particles can grow with time. 
It is conceivable that our compact astrophysical source may be in a highly mixed state because of the large interaction rate among its particles, the evolution of the state of a subset of the astrophysical system made of a finite number of neutrinos may also have a mixed state character. Yet, this fact does not directly imply that our subsystem is entangled. Assuming that our subset is actually entangled, as inferred from simplistic analyses (e.g.~see Ref.~\cite{Patwardhan:2022mxg} and references therein), it remains to be assessed whether such an entanglement may have physical consequences. In fact, as known in condensed matter physics, ``fluffy bunny entanglement'' could take place, i.e.~a kind of entanglement that is unavoidable, but useless and cannot be verified because of lack of access to the entire quantum state~\cite{doi:10.1126/science.1109545,Molmer}.

The arguments above highlight that in order to draw conclusions on the validity of the mean-field approach to investigate the flavor evolution in neutrino-dense sources, one needs to develop brand-new and physically motivated simulations. Given the nontrivial conceptual and technical challenges linked to the problem, in what follows, we focus on a simpler system (not aiming to mimic the physics of neutrinos within a dense source) and investigate the implications of some of the assumptions currently adopted in the neutrino many-body literature on the topic.

\section{Neutrino equations of motion in the many-body approach}
\label{sec3}
In this section, we briefly summarize the many-body formalism 
and begin by describing the initial state. We consider $N$ particles, which are either physical particles or systems that can be considered as a single quantum object. The initial wave function is constructed by taking the outer product of the flavor eigenstates. The state thus constructed has $2^N$ components in the two flavor approximation. 
Note that in this case the equation of motion has $2^{N}$ components as opposed to $2$ field equations foreseen within the  mean-field approach (see Appendix.~\ref{sec2}).
For example, if we consider a system consisting of two particles, the first one is $\nu_{e}$ and the second one is $\nu_{\mu}$, with two distinct momenta, the initial state in the many-body formalism is
\begin{eqnarray}
\begin{pmatrix}
|ee\rangle\cr
|e\mu\rangle\cr
|\mu e\rangle\cr
|\mu \mu \rangle\cr
\end{pmatrix}
=
\begin{pmatrix}
0\cr
1\cr
0\cr
0
\end{pmatrix}\ .
\label{inistate}
\end{eqnarray}
Since Eq.~\ref{inistate} is a wavefunction, it evolves in time according to the Schr\"{o}dinger equation. 
As the system evolves, the state described by Eq.~\ref{inistate} can become entangled and cannot be represented in the form of an outer product of two-component wavefunctions. 

The self-interaction Hamiltonian, which governs the evolution of the wavefunction, is derived by considering all possible momentum exchanges, including with itself. The $(i, j)$ entry of the self-interaction Hamiltonian is proportional to the number of ways in which momenta in the state $i$ can be exchanged to give the state $j$~\cite{Friedland:2003dv}. This is equivalent to keeping the momenta unchanged and counting the number of ways in which we can exchange the labels of the particles in our system.
For example, the state $|e \mu\rangle$ has four momentum exchanges possible; two with the exchange of momenta with itself, which retain the state, and two with the exchange of particles with the other particle, which convert the state to $|\mu e\rangle$. 
As a matter of convention, these numbers are divided by $2$ to take into account the double counting.
Thus, for our two-particle system in Eq.~\ref{inistate}, the self-interaction Hamiltonian is given by~\cite{Friedland:2003dv}:
\begin{eqnarray}
H \propto \begin{pmatrix}
2 & 0 & 0 & 0 \cr
0 & 1 & 1 & 0 \cr
0 & 1 & 1 & 0 \cr
0 & 0 & 0 & 2
\end{pmatrix}\ .
\label{hammb}
\end{eqnarray}
In order to ensure that the Hamiltonian has units of energy, the Hamiltonian is multiplied by $(\sqrt{2}G_{\textrm{F}}/V)(1-\cos\Theta)$, where $V$ is the normalization volume, $\Theta$ the angle between the momenta of the neutrinos, and  $G_{\textrm{F}}$ being the Fermi constant. 

The wavefunctions are uniformly distributed over the volume $V$, with periodic boundary conditions. 
The strength of self-interaction is 
\begin{eqnarray}
\mu_{0} \equiv \frac{\sqrt{2}G_{\textrm{F}}}{V}(1-\cos\Theta)\ .
\end{eqnarray}

 In the presence of the vacuum term, the flavor of each particle evolves independently of the other particles. Hence, in the  two-particle case, the vacuum Hamiltonian is given by,
\begin{eqnarray}
H_{\textrm{vac}} &=& \frac{\omega_{1}}{2}
\begin{pmatrix}
-\cos 2 \theta_{1} & \sin 2 \theta_{1} \cr
\sin 2 \theta_{1} & \cos 2 \theta_{1}
\end{pmatrix} \nonumber \\
&\otimes&
\frac{\omega_{2}}{2}
\begin{pmatrix}
-\cos 2 \theta_{2} & \sin 2 \theta_{2} \cr
\sin 2 \theta_{2} & \cos 2 \theta_{2}
\end{pmatrix}\ .
\end{eqnarray}
Here, $\omega_{i}$ and $\theta_{i}$ are the vacuum frequency and the vacuum mixing angle for the $i^{\textrm{th}}$ neutrinos. It should be noted that, if  time and $\omega_{i}$ are expressed in terms of $\mu_{0}$, the equation of motion is completely independent of the value of $\mu_{0}$ used.

\section{Plane waves in a closed interaction volume}
\label{sec4}

If there is a collection of $\nu_{e}$ and $\nu_{\mu}$ states, with momentum $\vec{p}$ and $\vec{k}$, respectively, then it is possible that two of these neutrinos will undergo momentum changing scattering. One possible outcome of the scattering is that we find a $\nu_{e}$ with momentum $\vec{k}$ and $\nu_{\mu}$ with momentum $\vec{p}$. From the perspective of a hypothetical observer, neutrinos with momentum $\vec{p}$, which were initially all $\nu_{e}$s, are now a collection of $\nu_{e}$ and $\nu_{\mu}$. Nevertheless, this scattering amplitude is not coherently enhanced by the number of neutrinos present, as we will see later.

\begin{figure}
\center
\includegraphics[width=0.49\textwidth]{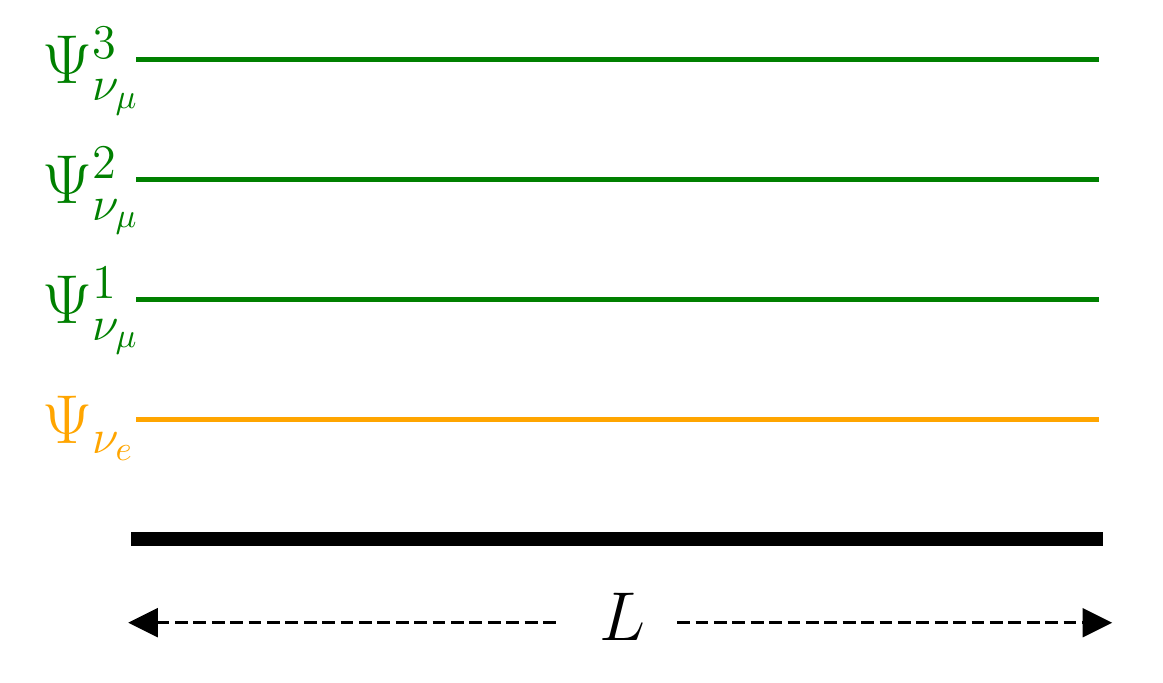}
\caption{Sketch of the setup of our system. The orange line indicates a plane wave of $\nu_{e}$s in the initial state, while the three green lines stand for three plane waves that are initially in the $\nu_{\mu}$ state (representative of the $1+3$ case of Fig.~\ref{Fig1}). The interaction volume is $V=L^{3}$, and periodic boundary conditions are assumed. Since the wavefunctions do not have spatial dependence, the neutrinos can be considered as stationary.  Since each plane wave is stationary, this implies that neutrinos interact with each other in the interaction volume for an indefinite period of time. The wavefunctions are shown in the cartoon as stacked one above the other for the purpose of legibility.
 }
\label{Fig0}
\end{figure}

We consider the temporal evolution of the neutrino ensemble sketched in Fig.~\ref{Fig0} made of two momentum beams, where 
a neutrino that is initially in the $\nu_{e}$ state (represented by a plane wave $\Psi_{\nu_e}$ that extends through the interaction volume $V=L^{3}$)
and undergoes interactions with a variable number of $\nu_{\mu}$s initially in the second beam (represented by $\Psi_{\nu_\mu}^{1, 2, 3}$ in Fig.~\ref{Fig0}). Throughout the evolution, the momenta of the neutrinos are restricted to one of the two beams.

\begin{figure*}
\includegraphics[width=0.49\textwidth]{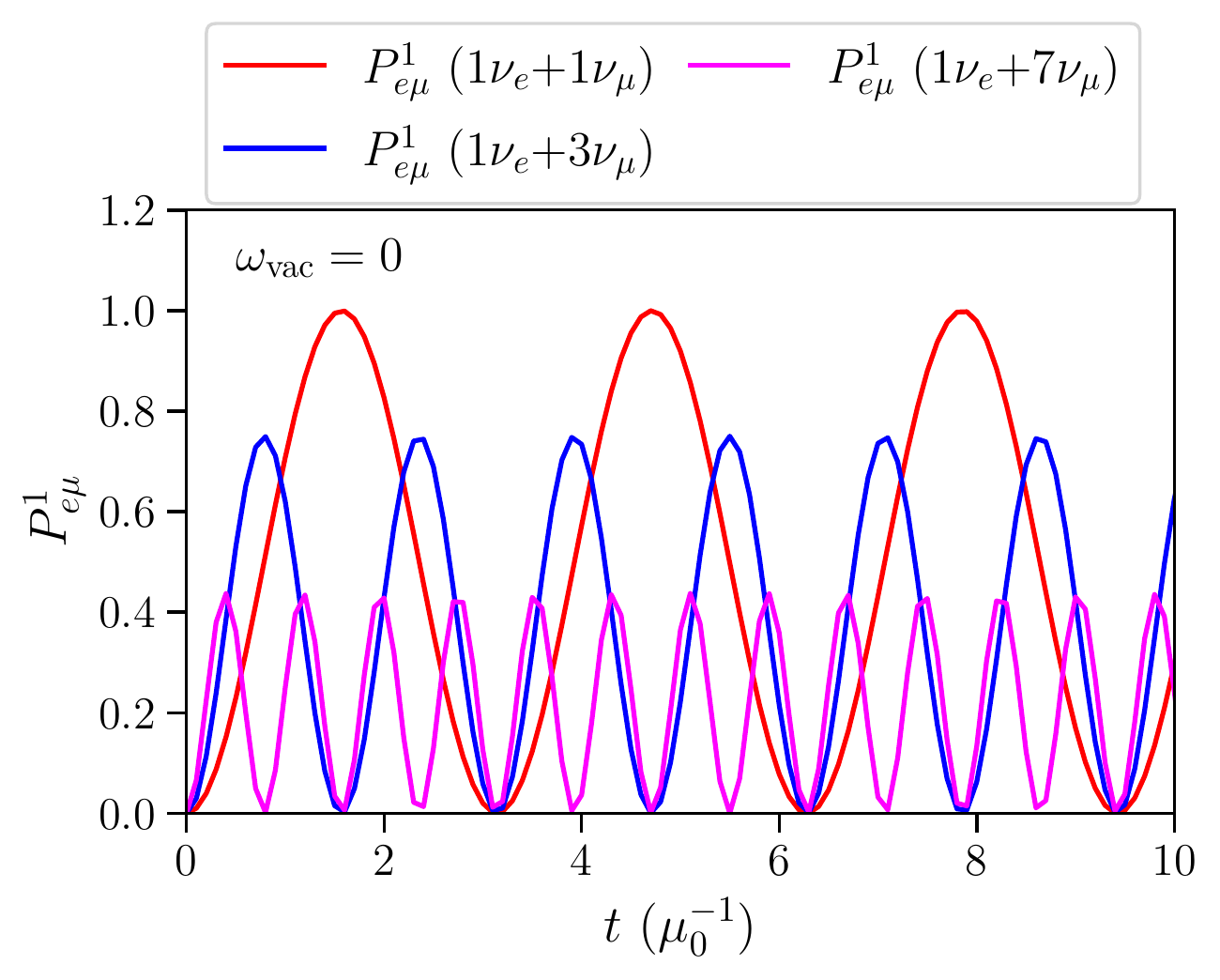}
\includegraphics[width=0.49\textwidth]{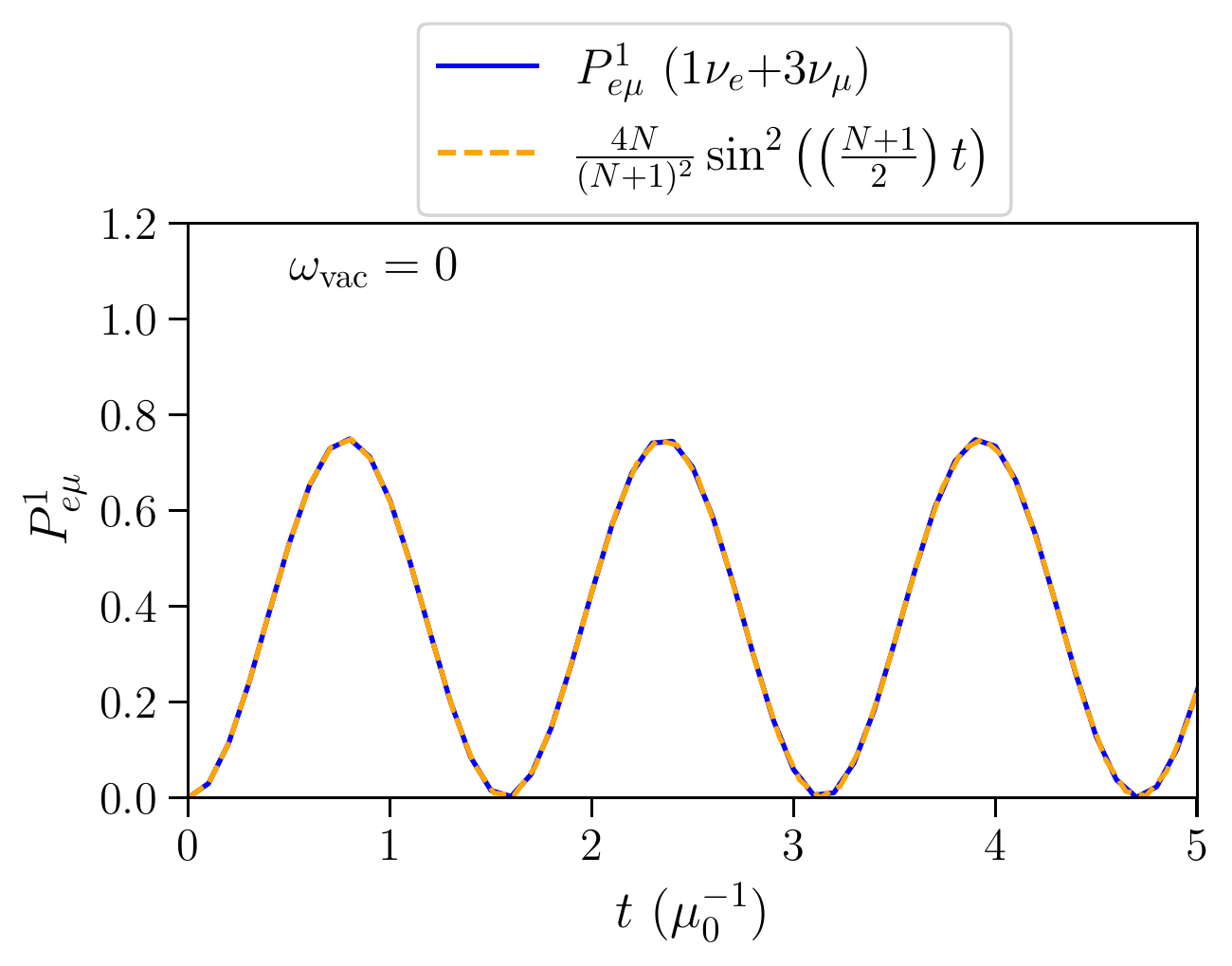}\\
\includegraphics[width=0.49\textwidth]{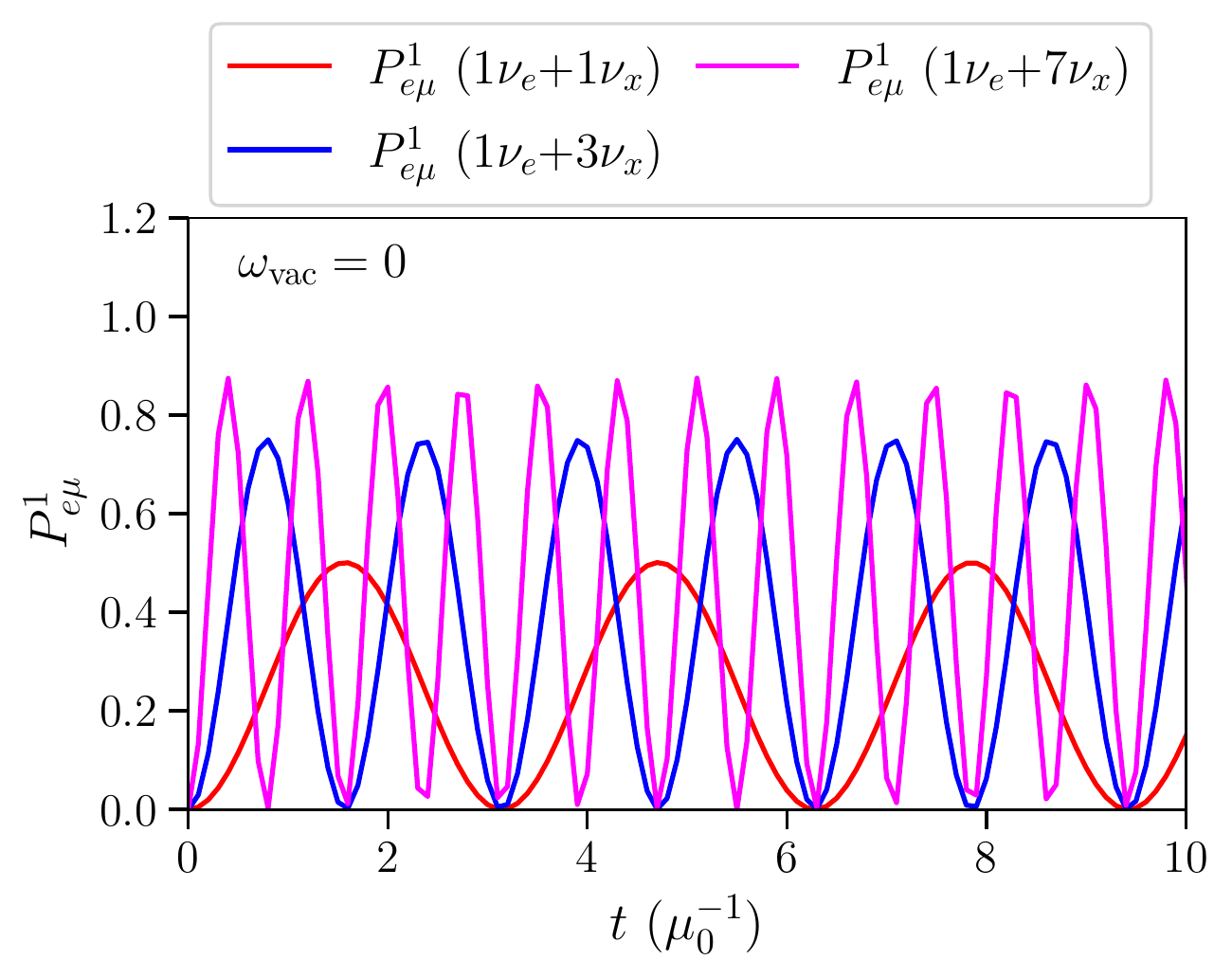}
\includegraphics[width=0.49\textwidth]{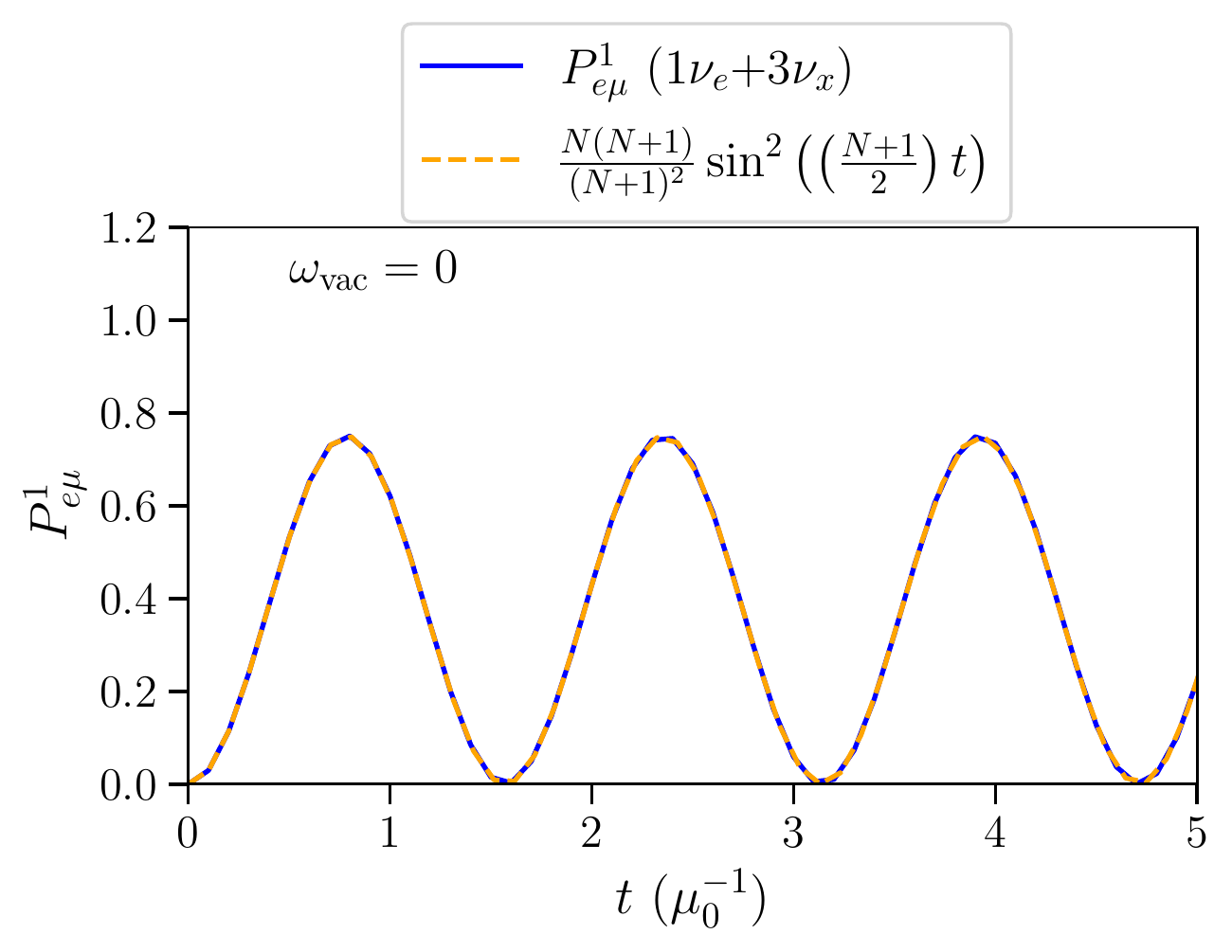}
\caption{Flavor transition probabilities for the system in Fig.~\ref{Fig0} and for one of the neutrinos being $\nu_{e}$ initially. {\it Top left:} Flavor transition probability of a neutrino that is initially in the $\nu_{e}$ state, interacting with N $\nu_{\nu}$ neutrinos for $\omega_{\textrm{vac}}=0$ and $N = 1, 3, 7$. The oscillatory behavior is solely due to the incoherent scattering term. {\it Top right:} Zoomed in version of the top left panel for the case with $3\ \nu_{\mu}$s in the background. The orange dashed line is the plot of Eq.~\ref{analytical1}, which is in perfect agreement with the numerical results.
{\it Bottom left:} Same as the top left panels, but for a neutrino that is initially in the $\nu_{e}$ state interacting with N $\nu_{x}$ neutrinos. 
In this case, the flavor evolution has a contribution from coherent forward scattering as well as incoherent scattering terms. 
{\it Bottom right:} 
Zoomed in version of the bottom left panel for the case with 3 $\nu_{x}$s (see Eq.~\ref{maxmix}) in the background. The orange dashed line represents Eq.~\ref{analytical2}.  
}
\label{Fig1}
\end{figure*}

In the mean-field formalism (see Eq.~\ref{eom1}) no flavor evolution occurs if collisions are ignored. The incoherent scattering leads to isotropization in the center of mass frame; so if incoherent collisions are included, we expect that both beams will have an equal number of each flavor of neutrinos. 
However, if we evolve the same system through Eq.~\ref{hammb}, flavor oscillations are seen as displayed in Fig.~\ref{Fig1}.

 For one $\nu_{\mu}$ initially in the second beam ($1\nu_e +1 \nu_\mu$ scenario), 
 we expect $50\%$ $\nu_{e}$s and $\nu_{\mu}$s in both beams. 
 This should be an equilibrium state, i.e.~no flavor evolution should be expected as a function of time when this configuration is reached. 
 If we increase the number of neutrinos in the second beam, e.g.~we have $1\ \nu_{e}$ in the first beam and $N\ \nu_{\mu}$s in the second beam, we expect that the equilibrium state of the first beam consists of a $1/(N+1)$ fraction of $\nu_{e}$, since in the equilibrium state both beams will have equal flavor content. (We have verified that this is the case numerically by solving the Boltzmann  equations by relying on the collision term only; results not shown here.

 When we solve the evolution equations for this system using Eq.~\ref{hammb}, we find that the flavor fraction in each beam does not reach equilibrium, as expected. Instead, we see an oscillatory pattern as shown in the top left panel of Fig.~\ref{Fig1}. This plot
 shows the temporal evolution of the flavor transition probability, $P^1_{e\mu}$, of the neutrino ($1$) that is initially in a pure $\nu_{e}$ state and undergoes interaction with a variable number of $\nu_{\mu}$s for $\omega_{\textrm{vac}}=0$. In order to obtain this plot and the following ones, we have used $\mu_0=10$~km$^{-1}$; however, since $t$ is expressed in units of $\mu_0^{-1}$, our results are independent of the value of $\mu_0$.

The reason for this peculiar behavior is due to the fact that when an incoherent scattering occurs, the amplitude for such a process is coherently added with the case of no scattering. 
\footnote{In scattering theory, when a particle scatters off a potential the wavefunction of the out-going particle is given by the sum of the incident (in) wavefunction and the scattered wavefunction:
$\Psi_{\textrm{out}} = \Psi_{\textrm{in}} + \Psi_{\textrm{scattered}}$.
The same holds in the system we consider, with  $\Psi_{\textrm{out}}$, $\Psi_{\textrm{in}}$ and $\Psi_{\textrm{scattered}}$ being plane waves. 
}
As a result, the neutrino, which was initially $\nu_e$, becomes a superposition of two flavor eigenstates. This can only happen if both the incoming and outgoing waves are plane waves.

We have verified that the dependence of the flavor evolution on the number $N$ of background neutrinos displayed in the top left panel of Fig.~\ref{Fig1} goes like $N t^{2}$ near $t=0$, in agreement with the findings presented in Ref.~\cite{Friedland:2003dv}. Coupling this finding together with the observation that the flavor evolution follows a sin-squared function, we find that the following empiric formula reproduces the results presented in the top left panel of Fig.~\ref{Fig1}:
\begin{eqnarray}
P^1_{e\mu}(t) = \frac{4N}{(N+1)^2}\sin^{2}\left[\left(\frac{N+1}{2}\right)t\right]\ ,
\label{analytical1}
\end{eqnarray}
where $t$ is in units of $\mu_{0}^{-1}$.
We demonstrate the excellent agreement between this empirical formula and the numerical results for $N=3$ in the top right panel of Fig.~\ref{Fig1}. Such degree of agreement holds for all cases with different $N$ for which we have numerical results.

 These results are consistent with the ones presented in Ref.~\cite{Rrapaj:2019pxz}, albeit the reason for the oscillatory behavior in the neutrino oscillation probability was not attributed to incoherent collisions in Ref.~\cite{Rrapaj:2019pxz}, rather to 
 differences between the many-body and the mean-field approaches. 
 However, this is not the case. The origin of the oscillatory behavior found in the many-body approach is actually due to the fact that incoherent scatterings are taken into account by construction in the many-body Hamiltonian (Eq.~\ref{hammb}), while only the coherent scatterings are considered in the mean-field approach (Eq.~\ref{Hamsn}).
 It is possible that, for the same reason, different outcomes for the flavor evolution in the many-body and mean-field approaches were found in Refs.~\cite{Cervia:2019res,Patwardhan:2021rej}.

The bottom panels of Fig.~\ref{Fig1} show the flavor evolution in the case of a $\nu_e$ scattering on a beam made out of maximal linear superposition $\nu_e$ and $\nu_\mu$:
\begin{eqnarray}
|\nu_{x} \rangle = \frac{|\nu_{e} \rangle+|\nu_{\mu} \rangle}{\sqrt{2}}\ .
\label{maxmix}
\end{eqnarray}
In this case, we expect flavor conversion in the mean-field approach in the absence of the collision term~\cite{Pantaleone:1992eq,Friedland:2003dv}. Due to the presence of both coherently enhanced scattering term as well as the incoherent collision term, the flavor evolution of $\nu_{e}$ due to $N~\nu_{x}$s in the background is expected to scale like ${1}/{4}(N^{2}+N)t^{2}$~\cite{Friedland:2003dv}, which is in agreement with our finding in the bottom left panel of Fig.~\ref{Fig1}. It should be noted that the period of oscillation in the top left plot and the bottom left plot is the same irrespective of the number of background neutrinos ($N$). Because of this, Eq.~\ref{analytical1} should be adapted to the case of a background consisting of $\nu_{x}$: 
\begin{eqnarray}
P_{e\mu}(t) = \frac{N(N+1)}{(N+1)^2}\sin^{2}\left[\left(\frac{N+1}{2}\right)t\right]\ .
\label{analytical2}
\end{eqnarray}
The bottom right panel of Fig.~\ref{Fig1} shows the agreement between Eq.~\ref{analytical2} and the numerical results for $N=3$. A similar formula can be obtained for a background that exists as any linear combination of flavor eigenstates. 

The oscillatory behavior seen in the left panels of Fig.~\ref{Fig1} can also be understood in terms of Rabi oscillations. The simplest case to understand is the $1+1$ case in which the flavor can be thought of as the orientation of a spin-$1/2$ particle; $\nu_{e}$ being the $S_{z}=1/2$ state and $\nu_{\mu}$ being the $S_{z}=-1/2$. The two particles can form a spin triplet state ($J=1$) or a spin singlet state ($J=0$). The initial state considered in our example is neither a triplet nor a singlet, but a superposition. Consequently, the probability of finding the system in a triplet or a singlet state evolves as a function of time, as shown in the top left panel of Fig.~\ref{Fig1}. As for the bottom left panel of Fig.~\ref{Fig1}, because of the composition of $\nu_{x}$ in terms of $\nu_e$ and $\nu_\mu$,
the combination of $\nu_{e}$ and $\nu_{\mu}$ for the singlet and triplet states is as described above, but the combination of two $\nu_{e}$s is purely a triplet state and not affected by Rabi oscillations. This results in a smaller oscillation amplitude in the $1+1$ case.

\begin{figure}
\includegraphics[width=0.48\textwidth]{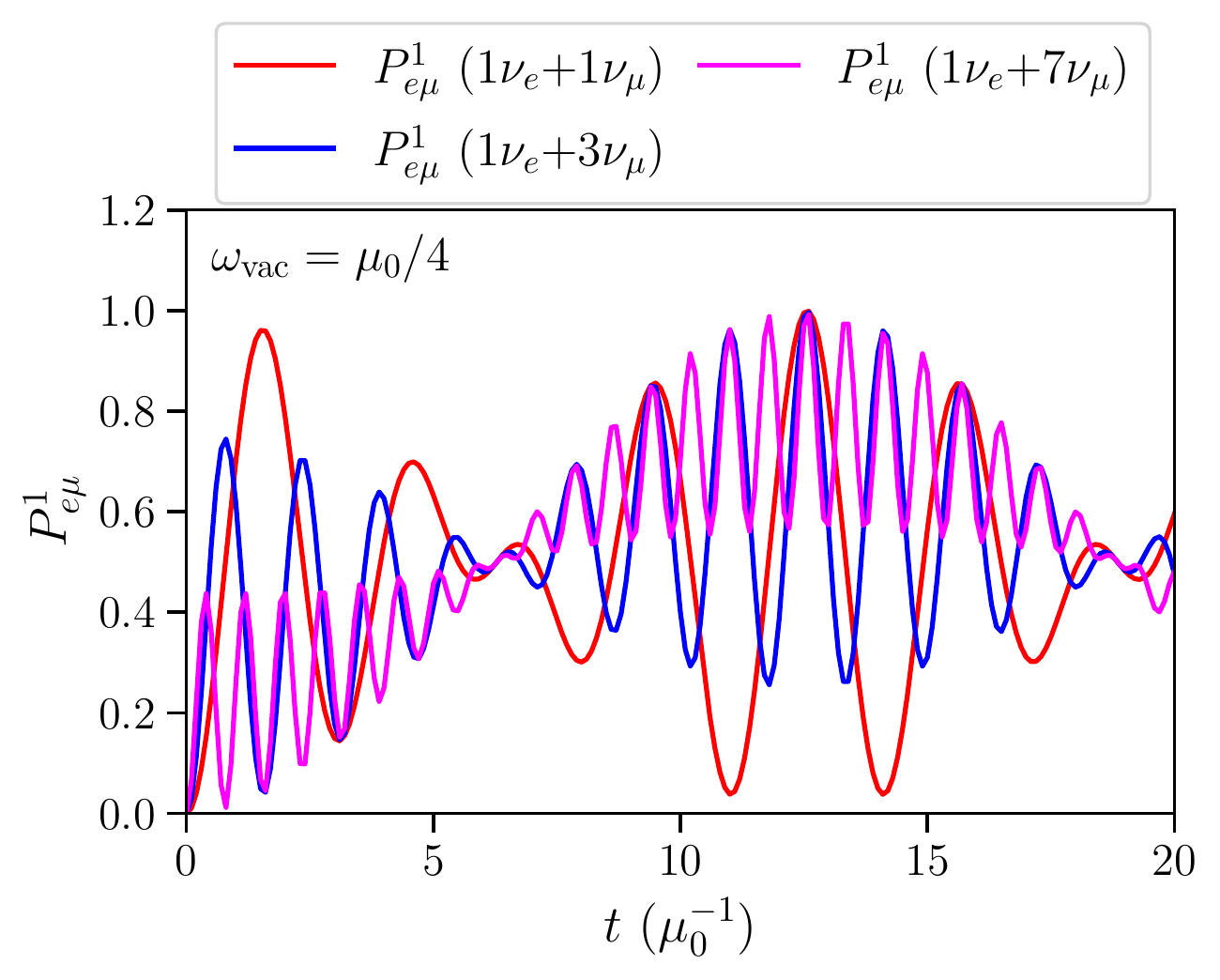}
\caption{Same as the top left panel of Fig.~\ref{Fig1}, but with $\omega_{\textrm{vac}} \neq 0$ and $\theta_{\textrm{vac}}=\pi/4$. The vacuum mixing term further modifies the oscillatory trend of the flavor transition probability. }
\label{Fig4}
\end{figure}
The results presented in Fig.~\ref{Fig1} can be further modified by the vacuum mixing term ($\omega_{\textrm{vac}} \neq 0$), as shown in Fig.~\ref{Fig4}, where maximal mixing angle $\theta_{\textrm{vac}}=\pi/4$ is assumed. 
We would like to emphasize that the change in the flavor evolution is gradual with increasing vacuum term. There are no sudden changes when the mixing angle is increased from zero to non-zero value as mentioned in Ref.~\cite{Rrapaj:2019pxz}. The same holds for increasing the vacuum frequency. For a system that is initially in the flavor eigenstates, the flavor evolution is solely due to the incoherent scattering term and increasing the vacuum frequency leads to a gradual change in the flavor evolution.

In this section, we have reported results obtained using the many-body approach for a finite number of background neutrinos $N$. These findings differ from the ones expected in the mean-field case. Yet, this does not prove that the many-body approach captures flavor conversion effects that are not obtained within the mean-field approach. In fact, the incoherent scattering term is proportional to $G_{\textrm{F}}^{2}N$, while the coherent forward scattering term is proportional to $G_{\textrm{F}}^{2}N^{2}$. Consequently, in the limit of $N \rightarrow \infty$, which should  mimic the mean-field scenario, the incoherent term should become negligible, and the result of the many-body formalism may be the same as the ones obtained by relying on the mean-field approach. 
 This ``expectation" was also reported in Ref.~\cite{Martin:2021bri}, but not demonstrated.
 Due to the limitations linked to the maximum value of $N$ that we can afford in our simulations, it remains to be assessed whether  incoherent collisions  may eventually lead to flavor conversion effects that are not captured by the mean-field equations. If any of such effects should be present, it will be crucial to assess whether they lead to such large changes on the flavor transition probability of neutrinos that could indeed have astrophysical implications. Note that, even if we were to perform many-body simulations with a large $N$, the impact of incoherent collisions will have to be evaluated moving beyond the plane-wave assumption, as discussed in the next section.

\section{Size of the wavepacket}
\label{sec5}

\begin{figure}
\includegraphics[width=0.49\textwidth]{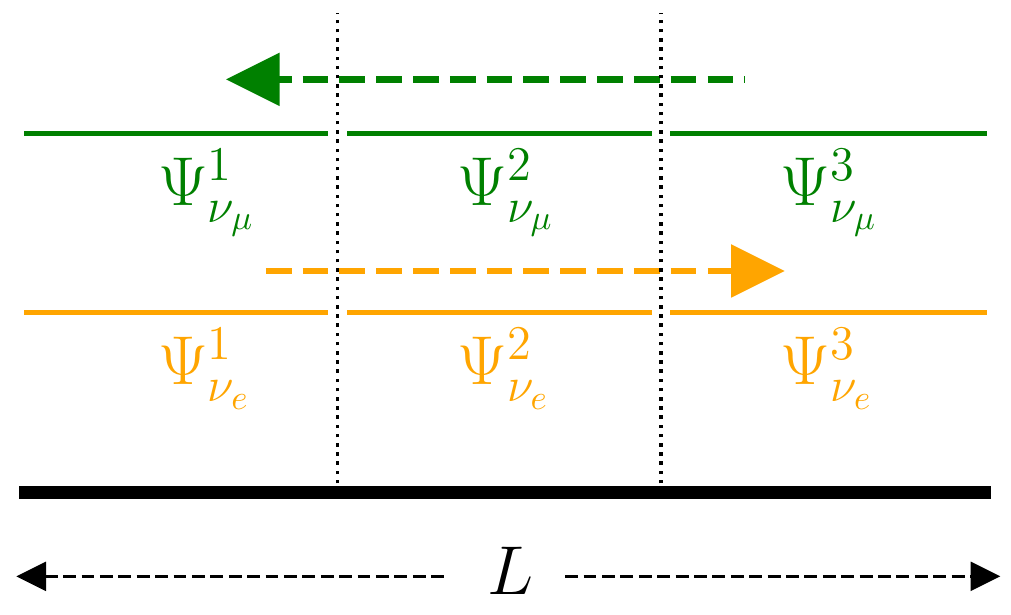}
\caption{Analogous of Fig.~\ref{Fig0}, but now with wavepackets of finite size instead than plane waves. 
The orange wavepackets show three neutrinos that are in $\nu_{e}$ state initially, and the three green wavepackets are neutrinos that are in the $\nu_{\mu}$ state initially. The arrows denote the direction of motion 
and periodic boundary conditions are assumed. The probability of interaction between the two neutrinos at a given time is expected to be proportional to the overlap between the two wavepackets. 
}
\label{Fig2}
\end{figure}
The assumption of infinitesimally small wavepacket size adopted in the mean-field approach is more suitable to mimic the physics of the interior of a neutrino-dense astrophysical source, which forms an open thermodynamic system. On the other hand, plane waves (adopted in Sec.~\ref{sec4} and more commonly in the literature on this subject) are rather motivated in the condensed matter systems where the particles are essentially stationary and form a closed system.

The results of Sec.~\ref{sec4} crucially depend on the assumption of using plane waves. To demonstrate this, we repeat the same calculation, however now using wavepackets. The geometry of our system is identical to the one introduced in Fig.~\ref{Fig0} with periodic boundary conditions, however, the neutrino wavepackets are now localized, each covering a third of the interaction volume, as shown in Fig.~\ref{Fig2}. 
The three orange wavepackets are initially $\nu_{e}$s, while the green wavepackets are $\nu_{\mu}$s. We assume that these wavepackets move in opposite directions with velocity $v$, as illustrated in Fig.~\ref{Fig2}. 

\begin{figure}
\includegraphics[width=0.49\textwidth]{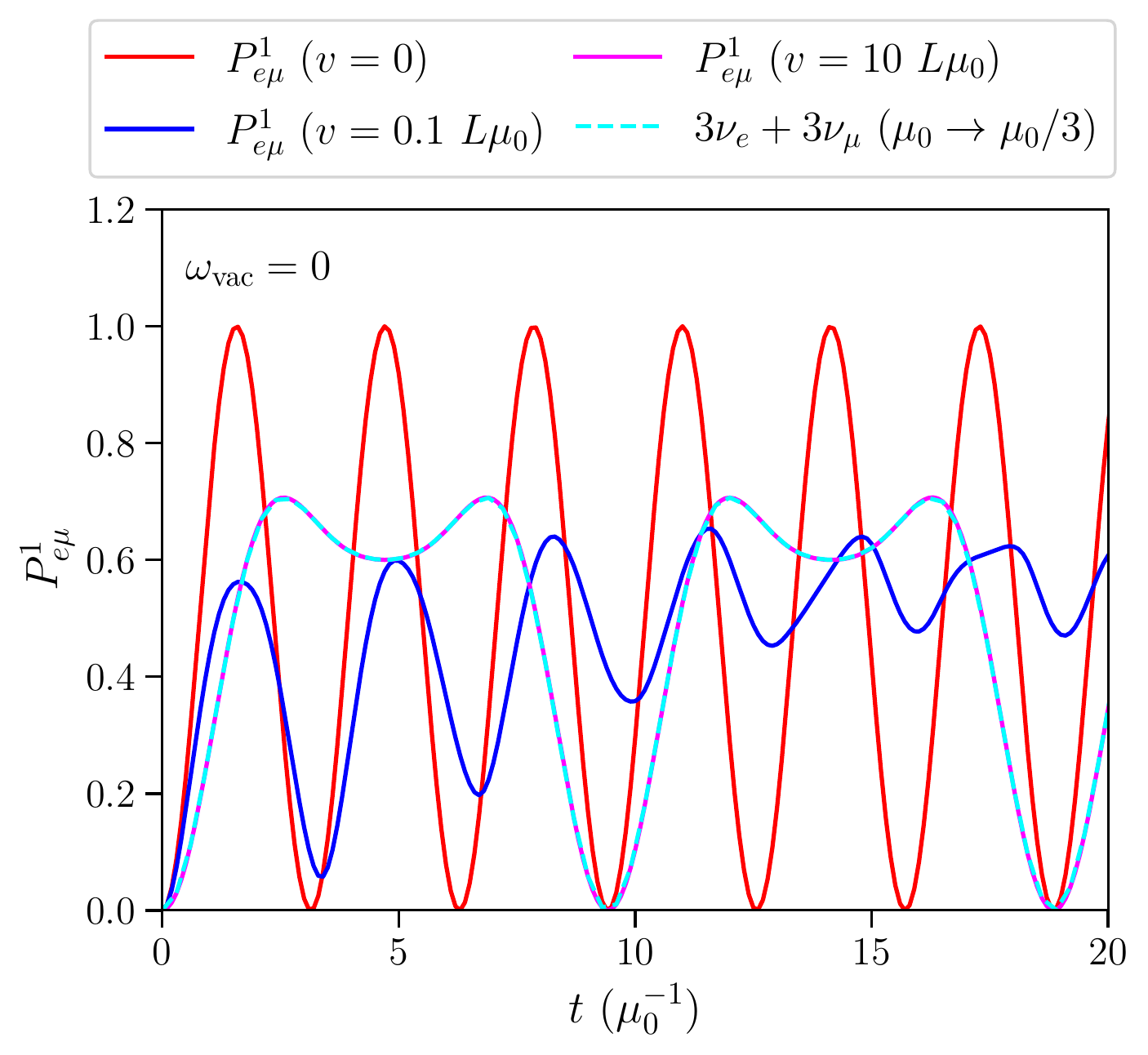}
\caption{Flavor transition probability for one of the neutrinos being $\nu_{e}$ initially and for the system sketched in Fig.~\ref{Fig2} as a function of time. We perform the simulation for different values of the velocity $v$ and $\omega_{\textrm{vac}}=0$. The flavor evolution is solely due to incoherent scatterings. 
}
\label{Fig3}
\end{figure}

If we assume that the velocity of the wavepackets is $v=0$, then each wavepacket interacts with only one other wavepacket of another flavor. The system behaves like the one used to obtain the results presented in Fig.~\ref{Fig1}.
The results of the many-body Hamiltonian are the same as the $(1+1)$ case shown in Fig.~\ref{Fig1}. This is indeed what we see in Fig.~\ref{Fig3}\footnote{It should be noted that the normalization for the wavepackets has been modified so that the normalization of the wavefunctions depicted in Fig.~\ref{Fig2} is same as the one of the wavefunctions depicted in Fig.~\ref{Fig0}; as a matter of convention we choose not to absorb this change of normalization in $\mu_{0}$.} for $v=0$, where we show the flavor transition probability of one of the neutrinos that starts as $\nu_{e}$.

The flavor evolution of the neutrinos is modified, if the neutrinos travel with non-zero velocity  (with the velocity being measured in units of $L\mu_{0}$), as shown in Fig.~\ref{Fig3}. As the neutrinos move, the interaction strength between the neutrinos changes, being proportional to the degree of overlap among wavepackets. 

As the velocity is increased,  which is equivalent to increasing the size of the box while keeping $\mu_0$ unchanged, the flavor evolution becomes independent of velocity (see cyan and magenta curves in Fig.~\ref{Fig3}) and converges to the case of three $\nu_{e}$s interacting with three $\nu_{\mu}$s with zero velocity, albeit with a self-interaction strength that is lower by a factor of $3$. This is not surprising since the neutrinos travel very fast in a system with periodic boundary conditions, and they see all the other neutrinos in a short period of time. However, this equivalence between these two systems is a consequence of the periodic boundary conditions. If we consider a train of wavepackets, such that each neutrino sees another neutrino for a brief period of time only, we should expect flavor equilibration. 
This kind of evolution would be expected in an astrophysical system where the typical time evolution period is much larger than the size of the neutrinos wavepackets, as also postulated in Ref.~\cite{Friedland:2003eh}.

\begin{figure}
\includegraphics[width=0.49\textwidth]{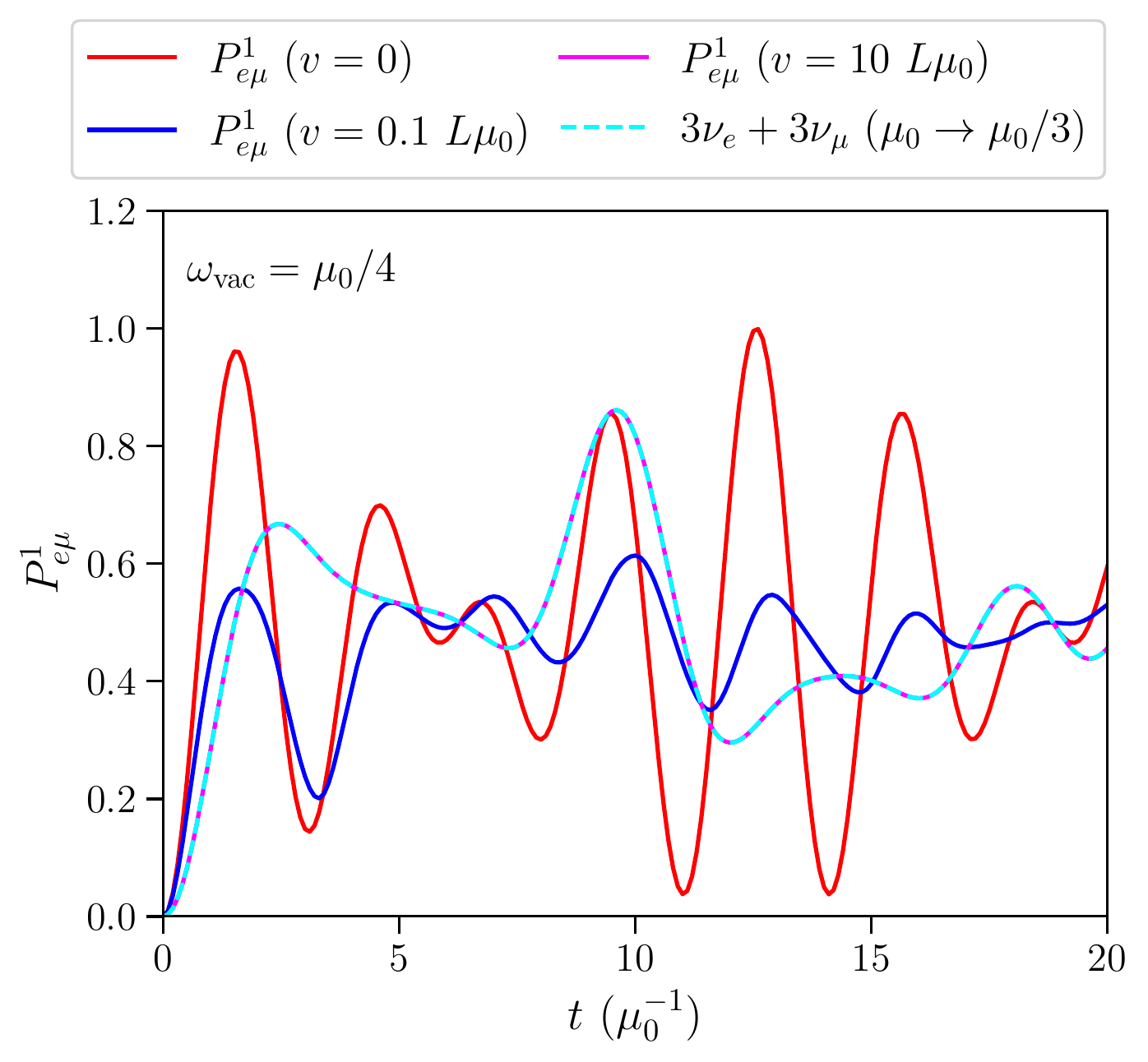}
\caption{Same as Fig.~\ref{Fig3}, but with $\omega_{\textrm{vac}} \neq 0$ and $\theta_{\textrm{vac}}=\pi/4$. Similar to Fig.~\ref{Fig4}, the vacuum term in the Hamiltonian further modifies the flavor transition probability in the many-body approach.}
\label{Fig4a}
\end{figure}
The results presented in Fig~\ref{Fig3} can be further modified by the vacuum mixing term ($\omega_{\textrm{vac}} \neq 0$), as shown in Fig.~\ref{Fig4a}. The comparison between Figs.~\ref{Fig4} and~\ref{Fig4a} suggests that the interplay between the incoherent scattering term and the vacuum term leads to a different flavor outcome using the plane wave approximation with respect to the wavepacket case. This also suggests that the plane wave assumption with static waves adopted in the literature is not reliable for gauging the flavor conversion phenomenology in a neutrino-dense source.

\section{Discussion and conclusions}
\label{sec7}
A growing body of literature aims to model neutrino flavor conversion in the presence of neutrino-neutrino interactions within the many-body formalism, through simple setups involving a small number of neutrinos [$\mathcal{O}(10$--$100)$ particles]. 
In this paper, we show that the different flavor outcome pointed out in the literature with respect to the widely used mean-field approach is due to the fact that the many-body and mean-field Hamiltonians are intrinsically different for a system containing a small number of neutrinos. In fact, a key feature of the many-body approach is that the momentum-changing incoherent scattering between the neutrinos is inherently present in the Hamiltonian. The incoherent collisions are not included in the mean-field Hamiltonian that by definition only involves the coherent forward scatterings of neutrinos. This is also responsible for the lack of entanglement entropy in the mean-field approach in the absence of non-forward collisions and its presence in the many-body context.  It remains to be assessed whether the entanglement found for many-body neutrino simulations with a small number of particles has any physical implications or it is of the fluffy-bunny kind.

Additionally, the small number of particles usually adopted in the many-body papers, as a consequence of technical limitations, prevents us from mimicking a system with an infinitely large number of particles that would resemble an ensemble for which the mean-field approach could make sense. Since these few particles (mimicked by extended plane waves) are constrained to interact for a long time within a closed volume with periodic boundary conditions, such a system has no similarities with the situation occurring in an astrophysical environment where an infinite amount of neutrinos (mimicked by wave packets of finite size) would stream in and out of a box in the absence of periodic boundaries. 

More importantly, we show a strong dependence of the flavor conversion phenomenology in the many-body approach on the
size of the wavepackets or equivalently their velocities. 
Also, the incoherent scattering term of the many-body approach cannot be replicated using the mean-field approach. 

On the basis of these findings, we conclude that the existing literature, focusing on the phenomenology of flavor conversion within systems of finite size and with a small number of particles in the many-body approach, cannot assess the validity of the mean-field approach to simulate the flavor evolution within an astrophysical system, neither it can mimic the behavior of neutrinos in the core of a supernova streaming out of the core.

Many open questions remain to be addressed concerning our understanding of the behavior of neutrinos in dense media. While the many-body approach might encapsulate features of neutrino interactions that might not be captured by a mean-field treatment, it is not clear whether the mean-field neutrino equations of motion could be corrected to include effects beyond the ones taken into account through the mean-field approach, analogously to the Bogoliubov de Gennes equation developed for Bose-Einstein condensates. 
 While preliminary formal work exists regarding higher order corrections that can be included in the mean-field approach to include two particle correlations using the Bogoliubov–Born–Green–Kirkwood–Yvon (BBGKY) hierarchy~\cite{Volpe:2013uxl, Froustey:2020mcq},  it is not clear whether and under which conditions the two particle correlations will become important.
We conclude that existing work cannot invalidate the mean-field treatment since it focuses on a system that does not reproduce the streaming flow of a large number of neutrinos simulated through the mean-field treatment in neutrino-dense astrophysical sources.

\acknowledgments
We thank Rasmus S.~L.~Hansen for involvement in the very early stages of this project. We are  grateful to  Klaus M\o lmer, Chris Pethick, and Georg Raffelt for many valuable discussions, as well as  Baha Balantekin, Rasmus S.~L.~Hansen, and Amol Patwardhan for comments on the manuscript. 
This project has received support from the Villum Foundation (Project No.~13164), the Danmarks Frie Forskningsfonds (Project No.~8049-00038B), and the Deutsche Forschungsgemeinschaft through Sonderforschungbereich
SFB~1258 ``Neutrinos and Dark Matter in Astro- and
Particle Physics'' (NDM).

\appendix
\section{Mean-field approach}
\label{sec2}

In this appendix, we summarize the neutrino equations of motion and related properties used within the mean-field formalism to model collective flavor conversion. 
For simplicity, we restrict the discussion to the two-flavor approximation without loss of generality. 

\subsection{Equations of motion} 
In the context of mean-field theory, the equations of motion 
can be written by relying on the density matrix formalism. In the two-flavor approximation, the density matrices for each momentum mode can be written as $2 \times 2$ matrices, $\rho$ and $\bar{\rho}$, for neutrinos and antineutrinos, respectively.

The diagonal components of the density matrix denote the occupation numbers for the flavor states, while the off-diagonal components contain information regarding coherence. Let us consider, e.g., the wavefunction which consists of a coherent superposition of $\nu_{e}$ and $\nu_{\mu}$: 
\begin{eqnarray}
\psi = \begin{pmatrix} a \cr b
\end{pmatrix}\ ;
\end{eqnarray}
the corresponding density matrix is 
\begin{eqnarray}
\rho = \begin{pmatrix}
|a|^{2} & ab^{*} \cr
a^{*}b & |b|^{2}
\end{pmatrix}\ .
\label{dencoherent}
\end{eqnarray}
The quantities $|a|^{2}$ and $|b|^{2}$ are proportional to the occupation numbers, which are conserved due to the unitarity of neutrino flavor evolution. If, on the other hand, we consider a mixture of $\nu_{e}$ and $\nu_{\mu}$, which are uncorrelated, e.g.~because the wavepackets are spatially separated and consequently without any coherence, then there are two separate wavefunctions,
\begin{eqnarray}
\psi_{\nu_{e}} = \begin{pmatrix}
a \cr 0
\end{pmatrix} \quad \mathrm{and} \quad
\psi_{\nu_{\mu}} = \begin{pmatrix}
0 \cr b
\end{pmatrix}\ .
\end{eqnarray}
The density matrix for such a system is given by 
\begin{eqnarray}
\rho = \begin{pmatrix}
|a|^{2} & 0 \cr
0 & |b|^{2}
\end{pmatrix}\ . 
\label{denincoherent}
\end{eqnarray}
Due to the absence of coherence between the two wavefunctions, the off-diagonal terms are absent. 

In the context of mean-field theory, it is assumed in numerical investigations that the initial state, or the starting point, is composed of a mixture of flavor eigenstates without any coherence between them. The implicit assumption is that all neutrino flavors are created independently in their flavor eigenstates. Hence, the density matrix is initially described by Eq.~\ref{denincoherent}. The off-diagonal components are then populated dynamically as the neutrino field evolves as a function of time.

In the mean-field approximation, the equations of motion that govern the flavor evolution  have the following form:
\begin{eqnarray}
i\left(\frac{\partial}{\partial t} + \vec{v}.\vec{\nabla}\right)\rho(\vec{x},\vec{p}) = [H(\vec{x},\vec{p}),\rho(\vec{x},\vec{p})]\ , \nonumber \\
i\left(\frac{\partial}{\partial t} + \vec{v}.\vec{\nabla}\right)\bar{\rho}(\vec{x},\vec{p}) = [\bar{H}(\vec{x},\vec{p}),\bar{\rho}(\vec{x},\vec{p})]\ ,
\label{eom1}
\end{eqnarray} 
where the barred quantities refer to antineutrinos. The density matrix for each spatial location $\vec{x}$ and momentum mode $\vec{p}$ evolves in accordance with Eqs.~\ref{eom1}, where $[\cdots,\cdots]$ on the right-hand-side denotes the commutator. The Hamiltonian for neutrinos or antineutrinos, $H$ or $\bar{H}$, contains three terms corresponding to the vacuum, matter, and self-interaction terms:
\begin{eqnarray}
\label{Hams1}
H(\vec{p}) = H_{\textrm{vac}} + H_{\textrm{V}} + H_{\nu\nu}(\vec{p})\ ; 
\end{eqnarray}
a similar expression holds for $\bar{H}(\vec{p})$, except for a minus sign preceeding $H_{\textrm{vac}}$. As for the terms in the Hamiltonian, 
they are defined as follows
\begin{eqnarray}
H_{\textrm{vac}} &=& 
\frac{\omega_{\textrm{vac}}}{2}\begin{pmatrix}
-\cos 2 \theta_{\textrm{vac}} & \sin 2 \theta_{\textrm{vac}} \cr
\sin 2 \theta_{\textrm{vac}} & \cos 2 \theta_{\textrm{vac}}
\end{pmatrix}\ ,\\
\label{Hammat}
H_{\textrm{mat}} &=& \begin{pmatrix}
\sqrt{2}G_{\textrm{F}} n_{e} & 0 \cr
0 & 0
\end{pmatrix}\ ,\\
H_{\nu\nu}(\vec{p}) &=& \sqrt{2} G_{\textrm{F}} \int (\rho(\vec{p^{\prime}})-\bar{\rho}(\vec{p^{\prime}})) \\
& \times &
(1-\vec{p} \cdot \vec{p^{\prime}}) d\vec{p^{\prime}}\ ;
\label{Hamsn}
\end{eqnarray}
the vacuum term is a function of the vacuum frequency $\omega_{\textrm{vac}}$ and vacuum mixing angle $\theta_{\textrm{vac}}$. The matter and self-interaction terms in the Hamiltonian result from the coherent forward scattering of neutrinos, with 
$n_{e}$ being the electron number density. The coherent forward scattering is a phenomenon due to which neutrinos undergo refraction in the medium when scattering off multiple targets, and the scattering amplitudes undergo constructive interference in the forward direction~\cite{Liu:1991ci, Langacker:1992xk}. 
In the case of a homogeneous gas, the spatial derivative in Eq.~\ref{eom1} can be neglected, as we assume in the rest of the paper. 

 It should be noted that in the presence of emission or absorption terms that may be present due to collisions of neutrinos with the background medium, the trace of the density matrix evolves dynamically;
 however, we ignore such terms~\cite{Shalgar:2020wcx,Hansen:2022xza}. In what follows, we briefly review the concept of coherent forward scattering before looking into the phenomenology that arises from Eq.~\ref{eom1}.

\subsection{Bogoliubov–Born–Green–Kirkwood–Yvon hierarchy}
The equations used in the many-body approach can be equivalently written as a series of equations using the BBGKY hierarchy~\cite{Volpe:2013uxl, Froustey:2020mcq}.
The equations of motion for neutrinos in the mean-field approach can be obtained from the first equation in the BBGKY hierarchy:
\begin{equation}
i\frac{d\rho_{1}}{dt} = [H_{0}(1),\rho_{1}] + \textrm{tr}_{2}[V(1,2),\rho_{12}]\ ,
\label{BBGKY1}
\end{equation}
where $\rho_{1}$ is the single-particle density matrix like the one in Eq.~\ref{dencoherent} and $\rho_{12}$ is the two-particle density matrix ($4 \times 4$ matrix). The subscript denotes the identity of the particles. The trace in the last term is the sum of the interactions of particle 1 with all possible particles 2. A similar equation can be written for antineutrinos. The two-particle density matrix is: 
\begin{eqnarray}
\rho_{12} \equiv \rho_{1}\otimes\rho_{2} + c_{12}\ .
\end{eqnarray}
If we ignore $c_{12}$, Eq.~\ref{BBGKY1} becomes
\begin{eqnarray}
i\frac{d\rho_{1}}{dt} = [H_{0}(1),\rho_{1}] + [\textrm{tr}_{2}(V(1,2)\rho_{2},\rho_{1}]\ .
\end{eqnarray}
This is the mean-field approximation. It should be noted that the term $c_{12}$ which encodes the correlations between the particles is responsible for incoherent scattering~\cite{Volpe:2013uxl, Froustey:2020mcq}. 
This implies that, in the mean-field approach, the incoherent scattering term is not present unless explicitly included. If the incoherent scattering term is included, then the equations cannot be technically called mean-field equations but are generally referred to as quantum kinetic equations (QKEs).
It is also possible to demonstrate that the mean-field equations can be derived from the many-body equations without using the BBGKY hierarchy~\cite{Balantekin:2006tg}.

\subsection{Coherent forward scattering}

In the case of non-forward collisions, neutrinos scatter off e.g.~electrons and change momentum. As for the coherent forward scattering, described by the matter term in Eq.~\ref{Hammat}, the interaction of neutrinos with a thermal bath of electrons is considered, and we need to take into account the scattering amplitude of neutrinos due to all the electrons in the interaction region. Since the electrons form a thermal bath, it is not possible to identify the electrons responsible for the scattering, and the scattering amplitudes due to scattering with all the electrons should be added coherently. The result is a constructive interference of the scattering amplitudes, if the scattered wave has the same momentum as the initial one and a destructive interference in all other cases. This phenomenon, where the scattering amplitudes from multiple scatterings is coherently added, is named coherent forward scattering. As a consequence of this process, the flavor conversion probability of neutrinos is modified~\cite{1978PhRvD..17.2369W, 1985YaFiz..42.1441M, Mikheev:1986if, Pantaleone:1992eq}.

It should be emphasized that the coherent sum of scattering amplitudes cannot be considered, if we identify the electron responsible for scattering. This is the same principle that is seen in a double slit experiment, where the interference pattern disappears, if we identify the slit through which the photon has passed~\cite{Feynman2009FeynmanVol3}. 

The same principles apply, if we consider the coherent forward scattering of neutrinos due to other neutrinos; however, additional considerations need to be incorporated. 
In fact, differently from the coherent forward scattering of neutrinos off electrons, the coherent forward scattering of neutrinos off other neutrinos has some amplitudes that preserve the flavor and others that do not preserve the flavor. The part of the amplitudes that preserves the flavor is identical to the case of neutrino coherent forward scattering from electrons. 
In addition, we need to consider amplitudes that result from neutrinos exchanging momentum, or equivalently flavor, with neutrinos in the thermal bath. Nevertheless, the flavor-changing amplitudes can only be added coherently sometimes. 

Let us consider an example of a neutrino scattering off a thermal bath of neutrinos that is comprised of an uncorrelated mixture of $\nu_{e}$s and $\nu_{\mu}$s. As discussed in Refs.~\cite{Friedland:2003dv,Friedland:2003eh}, since the neutrinos in the thermal bath are uncorrelated, the off-diagonal components of the density matrix are zero. Therefore, the flavor-changing coherent forward scattering amplitude is zero. The reason is that, if the incoming neutrino exchanges flavor with one of the neutrinos in the bath, the neutrino responsible for the scattering has also changed its flavor. Since we can identify the neutrino from the thermal bath responsible for the scattering, the scattering amplitude cannot be coherently added to another flavor-changing amplitude. Such a process is considered under the neutrino-neutrino collision term within the mean-field approach. It is generally not included in the calculations as it is very small compared to the refractive term. On the other hand, let us suppose that the neutrinos in the thermal bath are in a superposition of flavor eigenstates. In this case, knowing which neutrino from the thermal bath is responsible for the flavor exchange is not possible. This is why the flavor-changing amplitude is proportional to the off-diagonal component of the density matrices. Consequently, the resulting neutrinos are a coherent superposition of flavor eigenstates.

Note that we have ignored the momentum-changing processes that can also be present due to scattering between neutrinos. Although momentum-changing amplitudes are present, they do not contribute in the limit of large number densities in the mean-field approach. 
If we consider a dense neutrino gas that is spatially homogeneous, but not isotropic, and this gas is allowed to evolve as a function of time and in the absence of neutrino oscillations, we expect the neutrino gas to become isotropic due to incoherent scattering between neutrinos in the center of mass frame. In fact, it is possible to modify the neutrino flavor evolution equations (Eq.~\ref{eom1}) to include the collision term. 



\bibliography{notes.bib}
\end{document}